\documentclass[12pt, prd]{revtex4}
\usepackage{graphicx,epsfig}
\input{epsf.tex}
\usepackage{amsmath}
\usepackage{amssymb}

\usepackage[utf8]{inputenc}

\begin{document}

\title{The black hole fifty years after: Genesis of the name}
\author{Carlos A. R. Herdeiro
}
\affiliation{Departamento de F\'isica e CIDMA, Universidade de Aveiro,
Campus de Santiago, 3810-183 Aveiro, Portugal,
Electronic address:  herdeiro@ua.pt}

\author{Jos\'{e} P. S. Lemos}
\affiliation{Centro de Astrof\'isica e Gravita\c{c}\~ao - CENTRA,
Departamento de F\'isica, 
Instituto Superior T\'{e}cnico - IST,
Universidade de Lisboa - UL,
Avenida Rovisco Pais 1, 1049-001, Lisboa, Portugal,
Electronic address: joselemos@tecnico.ulisboa.pt}



\begin{abstract}
Black holes are extreme spacetime deformations where even light is
imprisoned. There is an extensive astrophysical evidence for the real
and abundant existence of these prisons of matter and light in the
Universe. Mathematically, black holes are described by solutions of
the field equations of the theory of general relativity, the first of
which was published in 1916 by Karl Schwarzschild. Another highly
relevant solution, representing a rotating black hole, was found by
Roy Kerr in 1963. It was only much after the publication of the
Schwarzschild solution, however, that the term black hole was employed
to describe these objects.  Who invented it? Conventional wisdom
attributes the origin of the term to the prominent North American
physicist John Wheeler who first adopted it in a general audience
article published in 1968. This, however, is just one side of a story
that begins two hundred years before in an Indian prison colloquially
known as the Black Hole of Calcutta. Robert Dicke, also a
distinguished physicist and colleague of Wheeler at Princeton
University, aware of the prison's tragedy began, around 1960, to
compare gravitationally completely collapsed stars to the black hole
of Calcutta. The whole account thus suggests reconsidering who indeed
coined the name black hole and commends acknowledging its definitive
birth to a partnership between Wheeler and Dicke.

\end{abstract}
\maketitle

\section{Introduction}
Einstein's general relativity, finalized in 1915, is a theory of
space, time and matter. Gravitation is explained by means of the
spacetime curvature and its interaction with matter. As John Wheeler
put it, spacetime tells matter how to move and matter tells spacetime
how to curve \cite{1w,2w}.

Black holes are objects predicted by general relativity. Since they
are made solely of curved space and time, and energy, they are the
most elementary objects of the theory. Stated in a simple form, black
holes are regions from which no physical body, not even light, can
escape. The boundary limiting this region is called the event
horizon. The spacetime deformation caused by a black hole challenges
our intuition. For instance, time flows arbitrarily slow in the
neighborhood of a black hole. Another example: for a rotating black
hole space is inexorably dragged by its rotation, making it impossible
to stay still, even outside of the black hole.

The first mathematically exact solution of Einstein's theory of
gravitation was the Schwarzschild solution \cite{3sch}, found in 1916
by the German astronomer and physicist Karl
Schwarzschild. Schwarzschild wanted to explain the gravitational field
outside a spherical star with this solution.  However, considered as a
complete solution without a central star, an intrinsic radius, called
the gravitational or Schwarzschild radius, pops out clearly from the
solution. This radius is very small relatively to the star's radius
and so it would not make part of the solution with a central
star. Moreover, the properties of the region it limits seem
odd. Consequently, this seemingly irrelevant region for astrophysical
considerations, which, additionally, is physically hard to grasp, was
for a long time largely ignored by physicists.

Yet, in 1939, a remarkable theoretical work by Oppenheimer and Snyder
\cite{4os} showed that the collapse of a star is complete, in its own
frame of reference, with the star and its surface passing through the
Schwarzschild radius without any resistance, continuing their
inexorable collapse. This paper was crucial towards a physical
understanding of the Schwarzschild solution that eventually surfaced
in the late 1950s, with the works of Wheeler, Kruskal and others
\cite{1w}. It then became clear that the gravitational radius yields a
surface named event horizon and that the region inside the event
horizon has interesting and peculiar properties.

The definite incentive to study these objects stemmed from the
detection, in the early 1960s, of radio sources of tremendous
energy. The additional discovery of an optical partner for these radio
sources led to the conclusion that these sources had a recession speed
of around 40\% of the speed of light, thus suggesting that they were
at cosmological distances, emitting a colossal amount of energy. These
sources became known as quasars, short for quasi-stellar objects,
because they looked like stars in the photographic plates
\cite{5c}. The release of massive amounts of energy led to the
speculation that highly compact objects, so compact that they could
even possess an event horizon, were at the source of these phenomena
and that relativistic processes were fundamental to generate these
energies.

A period of important new discoveries, both theoretical and
observational, followed. In 1963, Roy Kerr, a New Zealand physicist
working at the University of Austin, Texas, discovered the exact
solution for a vacuum rotating object in general relativity
\cite{6k}. If, in the Kerr solution, the rotation is put to zero, the
Schwarzschild solution is retrieved. The impact that both the
Schwarzschild and the Kerr solution had in physics and astrophysics
was such, that the solutions and the objects they represent needed to
be named. It was Wheeler that, in an article in 1968, named these
objects black holes \cite{7w}. The vacuum Schwarzschild and Kerr
solutions were thereafter called Schwarzschild and Kerr black holes,
respectively.

Concomitantly, astronomical observations over the last half-century
established the physical reality of these objects, that appear to
exist abundantly in the Universe. They result from the exhaustion of
fuel in massive stars, which then collapse under their own weight
\cite{8nm}. Additionally, supermassive black holes are known to exist
at the center of all, or almost all, galaxies and were formed in the
early stages of the Universe. Our own galaxy, the Milky Way, has a
supermassive black hole at its center, with about 4 million solar
masses \cite{9kh}. In 2015, the spectacular detection of gravitational
waves by the LIGO antennae \cite{10a} confirmed the existence of
black holes in a completely new manner, revealing a previously
inaccessible population.

It is interesting to speculate whether Einstein could have reached,
beyond the dark stars of Mitchell and Laplace, the black hole and
event horizon concepts back in 1905, solely based in Newton's
gravitation and the notion, from special relativity, that the speed of
light is both unsurpassable and a universal constant \cite{11j}.
Einstein, however, was never interested in either stars or in black
holes. In any case, the time was not ripe for this concept to appear
spontaneously, as history would show. Even after the theory of general
relativity being formulated and finalized, the path to reach the
concept of black holes and to grasp the significance of these objects
in their physical and mathematical plenitude would still prove
strenuous and convoluted \cite{12i,13j}.

The science of black holes is magnificent and the interest in these
objects has certainly entered the imagination and imaginary of human
culture. As such, it is only natural to ponder on the history of these
objects, starting\hskip0.06cmwith\hskip0.1cmthe\hskip0.1cmsimple\hskip0.1cmquestions\hskip0.06cmof\hskip0.06cmwhere\hskip0.06cmand\hskip0.06cmhow\hskip0.06cmthe\hskip0.06cmterm
black hole originated. As\hskip0.07cmwe\hskip0.07cmwill\hskip0.07cmsee,
there\hskip0.07cmis\hskip0.07cman\hskip0.07cmaura\hskip0.07cmof\hskip0.07cmmystery\hskip0.07cmand\hskip0.07cma\hskip0.07cmfascinating\hskip0.07cmstory.

\section{The initial names}\label{in}
The first physicists to dwell on the Schwarzschild solution believed
that at the Schwarzschild radius the gravitational field would be
infinite and thus spacetime appeared pathological in this
region. Hence, the Schwarzschild radius was also known as the
Schwarzschild singularity \cite{14r}. However, with the 1939 work of
Oppenheimer and Snyder it became clear that there was nothing singular
at that radius and so Schwarzschild singularity was an inadequate
terminology that did not faithfully represent the spacetime character
of the region.

In the 1960s, Soviet Union physicists, notably Yakov Zel'dovich and
Igor Novikov, put forth the term frozen star to describe this object
\cite{15z}.  This designation reflected the fact that for an external
far away observer, the star seems to be frozen when it reaches the
Schwarzschild radius, see \cite{16j} for a modern version of a frozen
star. On the other hand, American physicists, namely Wheeler, and
European physicists like Roger Penrose and others, applied the
terminology collapsed star \cite{17t}.  This name emphasized the
complete collapse of the star into a true curvature singularity in its
own frame.
	
But none of the names was a good name and Wheeler knew it. As to
correctly convey the physics and processes involved, a suitable name
is essential. In 1958, Wheeler proposed the term wormhole to describe
shapes that connected different regions, yielding spacetimes with
nontrivial topology. Ten years later he named collapsed objects as
black holes.

\section{John Wheeler's version for the term black hole}\label{BasicEq}

John Wheeler, in chapter 13 of his scientific autobiography Geons,
Black Holes \& Quantum Foam, with subtitle A Life in Physics
\cite{18w}, see Figure 1, explains how the term black hole was
introduced to describe the final state of a collapsing massive star.

\vskip 0.5cm
\begin{figure*}[ht]
\centering
\includegraphics[scale=0.57]{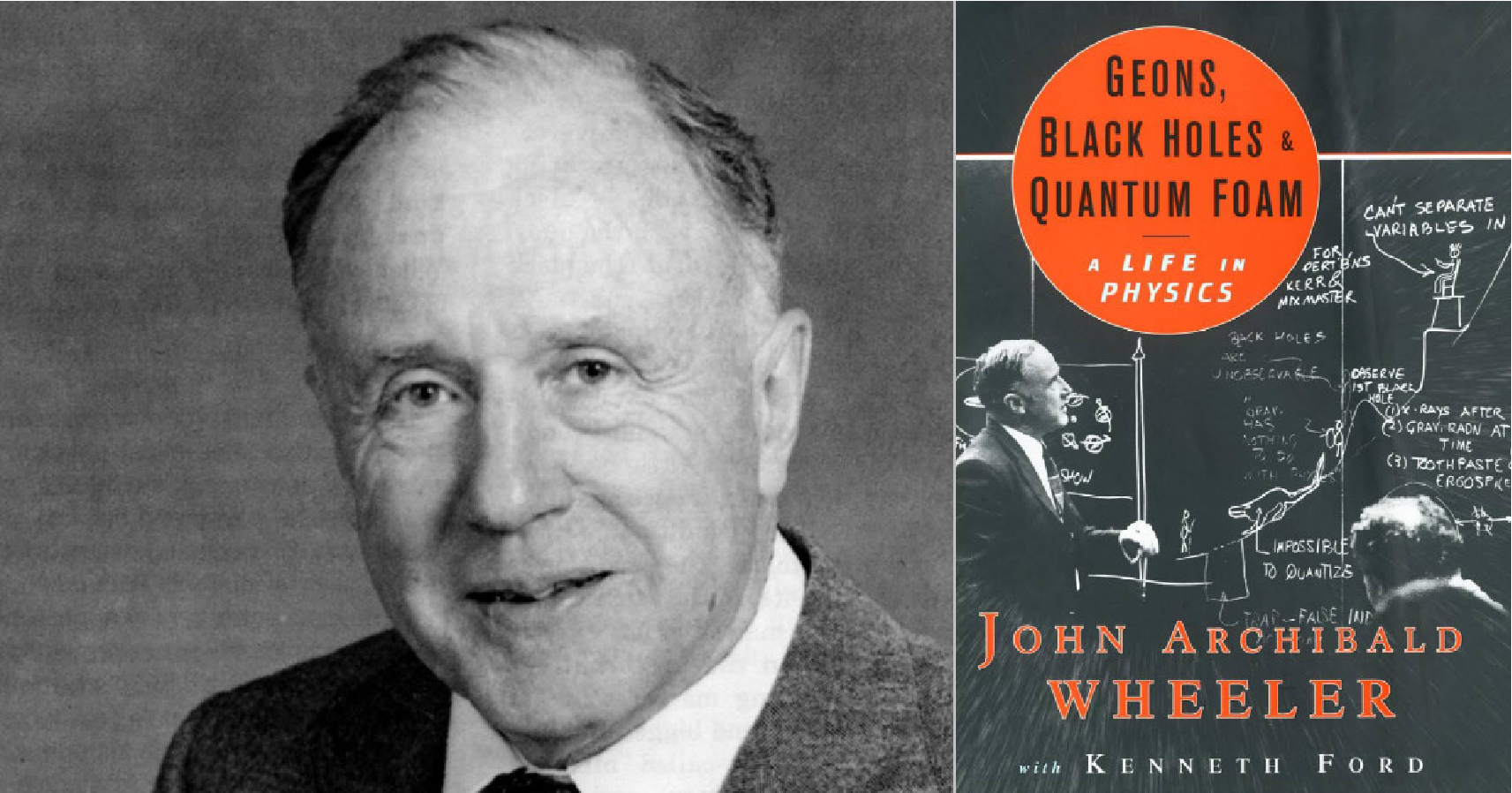}
\caption{ The physicist John Archibald Wheeler (1911-2008) and his
scientific autobiography \cite{18w}.}
\label{w}
\end{figure*}
\vskip 0.4cm

Wheeler, a physicist from the University of Princeton, excelled as a
nuclear physicist in his early career. He invented the S-matrix which
accounts for the quantum scattering of particle collisions and
participated in the Manhattan project that developed the first nuclear
weapon. In his autobiography, Wheeler describes how his interest in
nuclear physics and, subsequently in general relativity, led him in
the 1950s to turn his enthusiasm towards the study of stars and, more
specifically, to the stars' centers at the end of their lives. What is
the final state of a star after having exhausted all its nuclear fuel?
Does it explode? Or does it implode into a dense core of nuclear
matter? These were just some of the questions Wheeler wanted to find
the answer to. He was not the first to ask them and partial answers
already existed. It was known that the fate of a star depended on how
massive it was.

Low mass stars could end their lives as white dwarfs as suggested by
Subrahmanyan Chandrasekhar, an Indian physicist working at Cambridge,
around 1930. The Chandrasekhar limit is the maximum mass a white dwarf
can have, which is 1.4 times the mass of the sun. Following the ideas
of Fritz Zwicky, a Swiss astronomer settled at Caltech, there were
speculations that stars with higher masses could end their lives as
neutron stars. These stars had not yet been observed but were
theoretically predicted as highly compact bodies in which extreme
gravitational forces oblige the existing electrons and protons to
merge into neutrons. The star becomes a sort of giant atomic nucleus
and, as Landau, the renowned Russian physicist deduced, the value of a
neutron star mass limit was of the order of the Chandrasekhar
limit. Oppenheimer and Volkoff in 1939 confirmed Landau's prediction
for the neutron star maximum mass and they noted that it was
theoretically possible that nothing could stop the implosion of an
even more massive star. In turn, Oppenheimer and Snyder, still in
1939, through a simplified model, presented the famous result that a
star could collapse to the inside of its own gravitational radius
\cite{4os}. Later, in the 1950s, these problems were reconsidered by
Wheeler and his students. They showed that within a certain mass
interval, which they found to be slightly superior to the
Chandrasekhar limit, neutron stars could be stable and describe the
final state of the collapse of a star. However, it was the fate of the
most massive stars that intrigued Wheeler. After initial hesitations,
he ultimately accepted the generic validity of Oppenheimer and
Snyder's result.  For extremely massive stars their final state was
one of a gravitationally completely collapsed object (see
\cite{12i,13j} for a detailed explanation).

As Wheeler writes in his autobiography, after a decade considering
these ideas, in the fall of 1967, he was invited by Vittorio Canuto
from the NASA Goddard Institute in New York to attend a conference
about the nature of a new astronomical body, the pulsar. The pulsar
had been discovered a few months earlier by Jocelyn Bell Burnell and
Antony Hewish at Cambridge. It would soon be confirmed that pulsars
are rotating neutron stars.  In his lecture in New York, Wheeler
argued the possibility that at the center of a pulsar there could be a
``gravitationally completely collapsed object''. This terminology,
however, was long and inconvenient and Wheeler commented in his
lecture that he could not be repeating it all the time and a shorter
version was needed. At that moment, someone in the audience suggested,
``How about black hole?''. Wheeler writes that he found the term
perfectly appropriate for a ``gravitationally completely collapsed
object'', terminology which he had been searching for months. A few
weeks later, on December 29th 1967, Wheeler was invited to give a
talk, the Sigma Xi-Phi Beta Kappa lecture entitled ``Space and
Time'', at the annual meeting of the American Association for the
Advancement of Science in the New York Hilton \cite{19a}. There, he
employed the term black hole, which was then included in the written
version of the lecture, published in the spring of 1968 \cite{7w}. As
such, according to Wheeler, the name black hole entered the scientific
literature.

Bartusiak, a distinguished physics and astrophysics writer, confronts
Wheeler's memories \cite{20b}. Actually, the conference on pulsars and
neutron stars at the Goddard Institute, only took place in May 1968;
pulsars had been official announced in February of the same year. In
November 1967, there was in fact a conference about supernovas at the
Goddard Institute, but Wheeler's name is not in the proceedings of the
conference \cite{21b}. It is possible that he gave a lecture at this
meeting and did not send the written version to the editors; however
even this is not certain. What is irrefutable is that he uttered the term
black hole in the after-dinner speech at the annual meeting of the
American Association for the Advancement of Science and that the term
was printed in the American Scientist journal in 1968 in an article
entitled ``Our Universe: The Known and the Unknown''
\cite{7w}. Moreover, there is no doubt that with this publication the
name black hole was adopted worldwide and it became common usage
throughout all the sphere of knowledge, with Wheeler at its origin.

\section{The Texas Symposium in Dallas and the meeting of the
American Association for the Advancement of Science in Cleveland in 1963}

Irrespective of the details of Wheeler's tale on the origins of the
term black hole, there is a story he did not tell.

This story dates back to a scientific meeting that took place in
Dallas, in July 1963, the Texas Symposium on Relativistic
Astrophysics. The planning of this event had been motivated by the
recent discovery of quasars, the objects located at cosmological
distances that emitted colossal amounts of energy. In fact, part of
the scientific community began to speculate that relativistic
phenomena and concepts played an important role in explaining the
generation of these energies, sparking, in turn, the idea to hold a
conference on relativistic astrophysics \cite{22r,23s}, a
scientific field inexistent at the time.

Astoundingly, the term black hole was presented in the first Texas
Symposium, when discussing gravitationally completely collapsed
objects, a concept that scientists attending the meeting attempted to
relate with the huge amounts of energy emitted by quasars.

In the aftermath of the Symposium, Life magazine published an article
by Albert Rosenfeld, in the beginning of 1964, entitled ``Heavens' new
enigma, What are quasi-stellars?'' \cite{24r}. The article refers
to the Dallas encounter that had been held six months before and is
focused on quasars. It presents the idea, due to the astrophysicists
Fred Hoyle and William Fowler, that the energy source of quasars could
be linked to the gravitational collapse of matter, which in
Rosenfeld's words, would then result in an ``invisible black hole in
the universe'' \cite{24r}.

The term was maintained in another conference in Cleveland, in
December of 1963, promoted by the American Association for the
Advancement of Science. This conference prompted journalist Ann Ewing
to write an article named ``Black Holes in Space'' \cite{25e},
see Figure 2. The journalist starts the article with the statement
``Space may be peppered with black holes'', followed by ``Such a star
then forms a ``black hole'' in the universe''. This article is the
first time the term black hole appeared in print. Rosenfeld's article
on the Texas Symposium, held in July of the previous year, was
published with delay, six days after Ewing's article.

\vskip 0.4cm
\begin{figure*}[ht]
\centering
\includegraphics[scale=0.9]{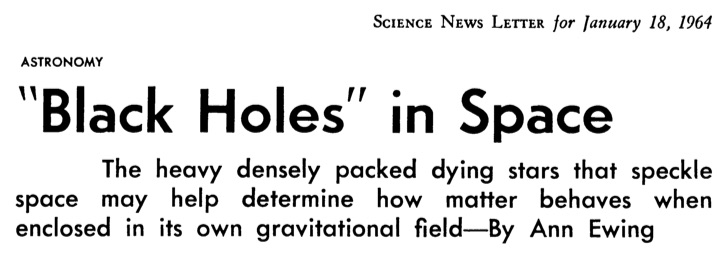}
\caption{Ann Ewing's article in 1964 where the term Black Hole
is published for the first time [25].}
\label{ae}
\end{figure*}

Both authors, Rosenfeld \cite{24r} and Ewing \cite{25e},
published the term black hole with the same meaning as Wheeler but
four years before Wheeler's article \cite{7w}. It is known that
they did not create the term. Marcia Bartusiak, states that Rosenfeld
himself said that he did not invent the term, he simply heard it
named in the Dallas meeting \cite{20b}. Notwithstanding,
neither Rosenfeld nor Ewing clarified which physicist or physicists
utilized the term in Dallas and Cleveland.

\section{Robert Dicke and the black hole of Calcutta }

The first Texas Symposium was so successful that it originated a
conference series, with the same name, held every two years. In the
twenty seventh Texas Symposium, in 2013, Bartusiak presented her paper
on the origin of the term black hole \cite{26b} (see also
\cite{20b}). Hong-Yee Chiu, an American astrophysicist of Chinese
origin that had invented the term quasar \cite{5c} and had actively
participated in the first Texas Symposium in Dallas, besides
organizing the meeting in Cleveland, confirmed that he had quoted the
name black hole in that meeting. In fact, the phrase ``Space may be
peppered with black holes'', with which Ewing starts off her article,
is attributed to him. However, Chiu denied having invented the term
black hole and traced its origin to a colloquium he was present in
1960 or 1961, as a postdoc, delivered by Robert Dicke, see Figure
3. At the colloquium, Dicke compared gravitationally completely
collapsed stars to the ``black hole of Calcutta''. According to Chiu
\cite{20b}, the term black hole was repeated by Dicke in his lectures
in 1961 and 1962 in the New York Goddard Institute, where Chiu now
worked. These lectures, that were part of a series of lectures given
also by other speakers including Wheeler, have been documented
\cite{27c} but the term black hole is not alluded to in the
proceedings.

\vskip 0.5cm
\begin{figure*}[ht]
\centering
\includegraphics[scale=0.74]{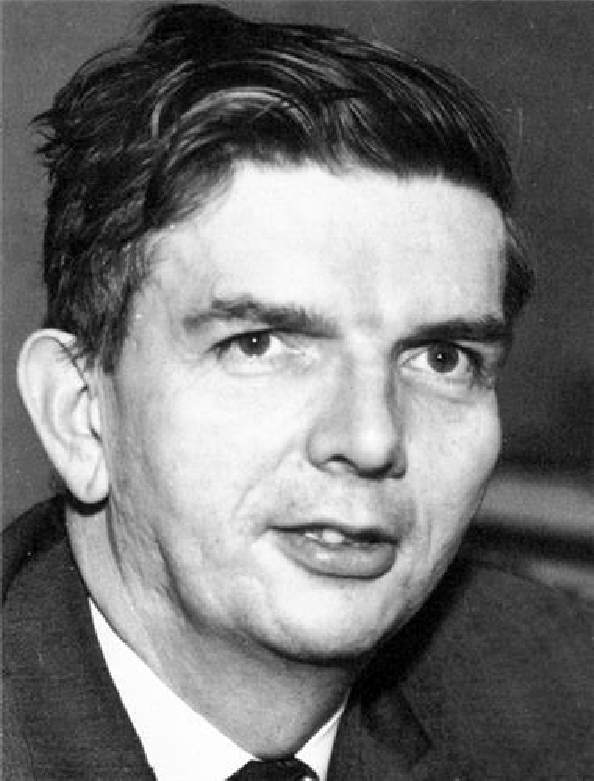}
\caption{ The physicist Robert Dicke (1916-1997).}
\label{d}
\end{figure*}
\vskip 0.4cm

Additionally, Martin McHugh, a physicist interested in Dicke's work,
told Bartusiak \cite{20b} a different and curious perspective on
the subject, see also \cite{28s} for other details. Robert
Dicke's children remember their father saying ``Ah, it must have been
sucked into the black hole of Calcutta'' whenever something got lost
at their house. What is the black hole of Calcutta Dicke is referring
to? There was a place named the black hole of Calcutta, sadly famous
in the history of British India. It was a small prison, in Calcutta's
Fort William, destined to no more than two or three prisoners at a
time. Following a dispute with the East India Company, which
controlled Fort William in the 18th century, the local governor Siraj
ud-Daulah, order a siege of the Fort, which was ultimately conquered
on June 20th, 1756. In John Howell's account of the episode, 146
soldiers of the East India Company were captured. The conquerors
confined the captives to the small prison in the Fort, known as
``black hole'' in the jargon of the soldiers. The cell was so
overcrowded that it was hard to shut the door. During that night, 123
of the 146 prisoners were suffocated or crushed to death. The details
of the incident are debated by other sources, but this version was
perpetuated during Britain's rule over India. There is even a monument
to the tragedy in St. John's church in Calcutta, in memory of those
who ``perished in the Black Hole prison'' \cite{29w}. This
account, of more than hundred men being crushed in a small space named
black hole, inspired Dicke to give that same name to an object
resulting from the total gravitational collapse of a star.

Thus, the term black hole appeared firstly through Robert Dicke.

\section{The plausible story behind the name black hole}

The term black hole was commonly mentioned since at least 1961 in
circles close to Wheeler. It thus seems odd why he omitted the
previous usages of the term in his account of the origin of the name
black hole, including the uses that appeared in print.

Canuto, the organizer of the meeting in the Goddard Institute in 1967
that actually took place in 1968, showed discomfort when confronted
with Wheeler's version presented in the book.  He even said, visibly
uncomfortable, ``Wheeler could have told the story he wanted''
\cite{30c}.

In fact, it is impossible that Wheeler and Dicke, as colleagues at
Princeton, never touched upon the subject or that Wheeler never heard
such an idiosyncratic expression as ``black hole of Calcutta'' in the
Princeton corridors.  Wheeler attended the lectures at the Goddard
Institute in 1961 and 1962, where Chiu declared hearing Dicke utter
the name black hole repeatedly.  He was also at the 1963 Texas
Symposium in Dallas, where the name black hole also appeared
recurrently, the article corresponding to his presentation on
gravitation theory and gravitational collapse appeared not in the
proceedings, but as a separate book with his collaborators Harrison,
Thorne, and Wakano \cite{22r}.  It seems that Wheeler did not go to
the meeting in Cleveland in that same year where the name was also
spoken, but all in all by 1963 he had been in at least two places
where the participants were talking freely about black holes already.

The name black hole did not fit well with some people. It was found to
be obscene, or even slang. Feynman, for instance, accused Wheeler of
being perverse for using this term \cite{20b}. For finding it
improper, the French physicists resisted the adoption of the name for
a long time, even when the name was already widely written and spoken
\cite{17t}.  Hence, it becomes clear that despite being accepted at
Princeton and amongst its sphere of influence in the early 1960s, the
term was not quite ready to be generally accepted. However, there was
a day where there was no more embarrassment and obstacles were put
aside forever.

What transpires is that at some point in the autumn of 1967,
presumably in New York, Wheeler decided to adopt the name
definitively, regardless of the other interpretations it might had,
and used the story that someone in the audience shouted the term at
him as a metaphor for another reality.

This interpretation is reinforced by José Ac\'acio de Barros, a
Brazilian physicist from San Francisco State University, that
collaborated with Patrick Suppes, a North American science
philosopher. According to Barros, in a conversation between him,
Suppes and Wheeler in 1996, Wheeler stated that the name black hole
was recurrent in his conversations with Dicke, and that there were
always playful smiles between the two when the name was brought up
\cite{31b}. Moreover, in that same conversation, according to Barros,
the person in the audience that screamed ``How about black hole?''
was Dicke himself \cite{31b,32j}.

And thus, the circle closes. History, when analyzed in detail, is
frequently richer that what it is portrayed by conventional wisdom. It
is intriguing that John Wheeler suppressed, in his description of the
origin of the term black hole, the former practices of this
terminology. In the history of the term black hole it is factual that
John Wheeler was neither the person who had the idea for the name nor
the first one to publish it with its modern scientific meaning. It was
Dicke that had that breakthrough and gave his consent to Wheeler by
shouting ``How about black hole?'', so that Wheeler could start using
the name freely.


\section{The acceptance, popularity and importance of
the name black hole}


When Wheeler in 1968 named as black hole a subsection of his article
\cite{7w}, he gave his authority to the term, and the name black hole
immediately captured the imagination of scientists and the public in
general. Throughout the following fifty years, from 1968 to the
present day, the name black hole was used an astronomical number of
times and now, fifty years after, it is clearly a date to be
celebrated \cite{33c}.

It really is the ideal name. The singularity creates a hole in
spacetime, preventing anything in the region within the event horizon
to escape out from it. For an outside observer, this hole is black,
since not even light emanates from it.

The 1916 Schwarzschild solution was initially considered as a solution
that described the exterior spacetime of a star, see \cite{34s} for
the celebration of the 100 years of the solution. However, when
considered as a vacuum solution, without a star, it presented problems
that physicists struggled to solve. Only several decades after the
solution was published and fully understood did it get its definitive
name: Schwarzschild black hole.  It is difficult to stress the extreme
importance of the Kerr black hole, i.e., the solution of a rotating
black hole. Astrophysical black holes have some degree of rotation,
small or large, and the Kerr black hole solution allows the study of
dynamic processes that mirror what is occurring in the neighborhood of
an observed black hole. Moreover, by bringing new dynamics to black
holes, the Kerr solution permitted new surprising theoretical
developments. For example, one was able to prove that ``black holes
have no hair'', another term created by Wheeler in 1971 to clarify
that black holes only had two properties: its mass and its angular
moment, see \cite{35h} for a modern discussion about this
theorem. Initially, this term also prompted some controversy. Despite
his serious sober temper, Wheeler had a naughty side to him, which was
shown and confirmed by Thorne when he wrote that, by coining the term
``black holes have no hair'', Wheeler generated a series of problems
with the editor of the Physical Review \cite{17t}, see also
\cite{36i}.

The extraordinary progress in black hole theory continued with the
possibility of energy extraction from rotating black holes via
superradiance and the Penrose process. It definitely culminated with
the discovery that black holes are thermodynamic bodies and with
Hawking's 1974 breakthrough that, via quantum processes, they emanate
radiation with black body temperature. But that is another story.

\newpage
\section*{Acknowledgements}

\vskip -0.3cm

We thank Luís Carlos Crispino for showing us the articles of Ann Ewing
and Tom Siegfried during the X Black Holes Workshop held in Aveiro in
December 18th and 19th, 2017, and for introducing us to the paper of
Alberto Saa, where these articles, that were in the origin of this
paper, are cited. We are grateful to José Acácio de Barros for sharing
with us his conversation with Patrick Suppes and John Wheeler.

We thank Teresa Sande Lemos Cunha e Sá for a first translation of the
paper from Portuguese to English.  The Portuguese version has been
published in Gazeta de Física {\bf 41(2)}, 2 (2018),
https://www.spf.pt/magazines/GFIS/392.

We acknowledge Funda\c c\~ao para a Ci\^encia e Tecnologia (FCT)
Portugal for financial support through Grant~No.~UID/FIS/00099/2013.
We are indebted to Science News for permission to utilize Figure 2.

\newpage

\end{document}